# Miscomputation in software: Learning to live with errors

Tomas Petricek[a]

a   The Alan Turing Institute, London, UK

**Abstract**   Computer programs do not always work as expected. In fact, ominous warnings about the desperate state of the software industry continue to be released with almost ritualistic regularity. In this paper, we look at the 60 years history of programming and at the different practical methods that software community developed to live with programming errors.

We do so by observing a class of students discussing different approaches to programming errors. While learning about the different methods for dealing with errors, we uncover basic assumptions that proponents of different paradigms follow. We learn about the mathematical attempt to eliminate errors through formal methods, scientific method based on testing, a way of building reliable systems through engineering methods, as well as an artistic approach to live coding that accepts errors as a creative inspiration.

This way, we can explore the differences and similarities among the different paradigms. By inviting proponents of different methods into a single discussion, we hope to open potential for new thinking about errors. When should we use which of the approaches? And what can software development learn from mathematics, science, engineering and art?

When programming or studying programming, we are often enclosed in small communities and we take our basic assumptions for granted. Through the discussion in this paper, we attempt to map the large and rich space of programming ideas and provide reference points for exploring, perhaps foreign, ideas that can challenge some of our assumptions.

## The Art, Science, and Engineering of Programming

## 1  The history and classification of errors

> *If trials of three or four simple cases have been made, and are found to agree with the results given by the engine, it is scarcely possible that there can be any error among the cards [5].*

The opening quote from Charles Babbage about the Analytical Engine suggests that Babbage did not see errors as a big problem. If that was the case, the software industry would save billions of dollars, but sadly, eradicating all software errors turned out to be harder than expected.

In retrospect, it is curious to see how long it took early computer personnel to realise that coding errors are a problem. One of the first people to realize that "[he] was going to spend a good deal of [his] time finding mistakes that [he] had made in [his] programs" was Maurice Wilkes in late 1940s [17]. However, as Mark Priestley writes, a typical comment [circa 1949] was that of Miller, who wrote that such errors, along with hardware faults, could be "expected, in time, to become infrequent" [41]. While hardware faults did become relatively infrequent, the same cannot be said about coding errors.

Over the next 20 years, computers developed from a research experiment into an ordinary scientific tool and industrial instrument. As noted by Nathan Ensmenger, by the end of 1960s many were talking of a computing crisis. For the next "several decades, managers, academics and governments would release warnings about the desperate state of the software industry with ritualistic regularity" [17].

In this essay, we trace the history of errors, or miscomputations [19], in software engineering. We look at different kinds of errors and discuss different strategies that programmers employ for dealing with errors, taking inspiration from mathematics, science, engineering and art. We conclude by speculating how can the industry escape from the seeming perpetual computing crisis.

**Lesson 1: Responding to the computing crisis**

Teacher: To open the discussion, do you think that software engineering is suffering from a crisis caused by coding errors? And if so, what do we need to eradicate them?

Pupil Beta: I'm not sure we can really call our discipline software *engineering*. As a proper engineering discipline, we need to develop engineering standards that ensure minimum performance, safety requirements and make sure that developing software is a consistent and repeatable process. The black art of programming has to make a way for the science of software engineering.[1]

Pupil Alpha: This is an admirable goal, but building software is not repeatable in the same way as, say, building bridges. In software, requirements are always changing. We should see software as craftsmanship and emphasize skills of developers.

---

[1] The quote is adapted from the calls for the professionalization of programmers that appeared around the time of the NATO Conference in 1968 [17]

The only way to make software better is to value individuals and interactions over processes and tools[2]!

Pupil Tau: Coding errors must be avoided at all costs. If a program contains an error, we should not even call it a program! But you are both going about it the wrong way – even with the best craftsmans and processes, there is always a chance of human error.

We need to build on mathematical foundations and write provably correct software. Building on strong theoretical foundations, ideas like dependently typed programming are the way toward eradicating all errors. Striving for anything less is unethical!

Pupil Alpha: This might work for the factorial function or Fibonacci numbers, but real software systems are not so simple. When you start building a real system, you do not even have a full specification, let alone a formal one.

What really matters is how we discover the specification and this is where test-driven methodology can help [7]. You turn requirements into tests, discovering the specification and writing software that adheres to it at the same time.

Pupil Epsilon: Oh, come on! You both really think you can eliminate all errors? Take any complex distributed system as an example. So many things can go wrong – one of your machines runs out of memory, your server receives an unexpected request. Even a cosmic ray may flip a bit in your memory if your data centre is large enough! The path to dependable software is to learn to live with errors.

Pupil Tau: I would like to object; we can accommodate for all of these situations in our proofs, but I see there is only little time left and I'm curious if Omega can direct our discussion back into a more sensible direction.

Pupil Omega: Yo, I tell you, errors are fun!

Teacher: That is an interesting position Omega, but I think the class needs more explanation. What exactly do you mean?

Pupil Omega: I was playing in a club yesterday. I accidentally put on a wrong sample and it turned out to be much better. If you make an error, you might be surprised. And if Tau thinks it is unethical, wouldn't it be just as unethical as to limit our method of discovery? Penicillin was discovered by the kind of accident that Tau wants to ban!

Teacher: It does seem that after more than 60 years, errors are still an important concern of the software industry and there is much disagreement about how to live with them. Perhaps it will be easier to expand on the different positions once we consider recent infamous examples of concrete coding errors.

**Lesson 2: From Y2K to the Knight Capital glitch**

Teacher: Let us now consider the different strategies for dealing with errors using a number of case studies. The first one is the infamous Y2K bug. This was caused by the fact that many old programs stored only the last two digits of a year in a date and so adding one day to 31/12/1999 produced a date that could be interpreted as 1/1/2000, but also as 1/1/1900.[3]

---

[2] Alpha paraphrases the position of the craftsmanship movement [33], which has been linked with art and contrasted with the engineering attitude [43].

As a concrete instance, my bank statement on 1/1/2000 reported that $8617 disappeared from my savings! Fortunately, they returned my money on the next day...

OMEGA: Ouch! Despite all the Y2K craze,[4] I had the impression that the bug caused only relatively small number of problems. It is an interesting case where the potential issue was discovered before it actually occurred. Yet, it was surprisingly hard to fix.

BETA: I can see how the bug in the banking system could have happened. I bet the specification was correct and stated how to calculate the compound interest for a given number of years, but due to a sloppy encoding of years, the system calculated the interest based not on +1 year, but based on -99 years.[5]

OMEGA: This is a case where formal proofs are not going to help. You can prove that your calculations and algorithms are correct, but when it comes down to metal, years won't be idealized natural numbers, but only two digits to save memory space!

TAU: That would only be the case if you were using program proofs in an informal way, but if we talk about proving programs correct, we must not make any idealizing simplifications. We must prove properties of the actual implementation – the bug was an incompleteness in the specification. The specification must be detailed enough not to leave room for any ambiguity, such as how are the dates going to be represented.

ALPHA: I cannot easily think of a theorem to prove here. If the property we prove is that the system correctly calculates the return on investment using the compound interest formula for a given number of years, that still does not prevent us from accidentally calculating it for -99 years...

TEACHER: Finding properties to prove is an important question. We will return to it in Lesson 6. To find further interesting questions to discuss, let's consider another well-known bug. In August 2012, the trading firm Knight Capital lost $440m in less than 45 minutes due to an erroneous deployment of their software.

The company removed outdated code from the software and re-purposed a flag that was used to trigger it. It then deployed a new version of the software on all servers except for one – and turned on the flag. This executed the new routine on the updated servers, but it triggered the outdated code on the one server running the old version, causing the system to issue millions of erroneous trades [46].

OMEGA: I cannot believe it took Knight Capital 45 minutes to turn off the malfunctioning server! The lesson here is clear. You should design the system to make such errors immediately visible and you should be able to quickly manually intervene. You should be able to stop incorrect trading just like you can stop a dissonant chord that you accidentally play in live music performance.

---

[3] Unlike a modern data scientist, our class is lucky enough that it does not need to consider different ways of writing days and months in a date!
[4] How media reported on the Y2K bug has been well-documented by Davies [14]
[5] BETA makes a correct guess, $8617 is what you lose when you assume 2% interest rate on initial investment $10000 over -99 years.

Epsilon: I can see how a manual coding intervention could correct the error promptly, but why not avoid it in the first place? The scenario sounds exactly like the situation for which Erlang and its "let it crash" [3] approach has been designed! Erlang has been used in telecommunications for systems that need to run for years without an outage and are updated on the fly.

Rather than re-purposing a flag, you would simply add a new kind of message. If an old system was running on one of the servers, it would crash when it receives a message it cannot handle – and it would get restarted by a supervisor process while all the updated servers would continue working fine.

Tau: If the system is programmed to automatically recover, then I do not think we are still talking about errors. But I must admit, I find it a very inelegant approach...

Teacher: The case of Knight Capital revealed two more important questions that we should discuss in future lessons. First, I would like to return to the idea of live intervention in Lesson 5. Second, the Erlang approach of incorporating some degree of error tolerance is an interesting alternative and we'll return to it in Lesson 10.

Now, let's discuss one more case study. In January 2016, Google Translate caused an embarrassing diplomatic incident. When translating from Ukrainian to Russian, it started translating Russia as "Mordor" and Sergey Lavrov (Russia's foreign minister) as "sad little horse" [6]. The error was introduced automatically – it mirrored language used by some Ukrainians following Moscow's annexation of Crimea in 2014.

Tau: I object, this is not a real error!

**Lesson 3: Classifying kinds of miscomputation**

Teacher: We uncovered an interesting issue last time – is the unexpected behaviour of Google Translate actually an error? To answer this, perhaps we can start by trying to understand what kinds of errors or *miscomptuations* [19] are there.

Beta: Let me try. First, there are simple *slips* where your idea was right and you tried to encode it in the right way, but you made a syntax error, a typo or, say, a copy-and-paste error. Then there are *failures* where you had the right idea but encoded (some part of) it poorly. Finally, *mistakes* are the kinds of errors where you made an error when thinking through the specification.[6]

Tau: Knight Capital glitch would then be a failure, while Y2K bug is a mistake because nobody expected software to still run in the year 2000. However, Google Translate simply followed the specification of its machine learning algorithms...

Alpha: That might be the case if the specification said "run this deep neural network learning algorithm on data scraped from Ukrainian internet", but I doubt that is the case. We must see programs in the wider socio-technological perspective.[7] You wouldn't want an accident like that to cause a war...

---

[6] Beta is paraphrasing categorization of miscomputation by Primiero and Fresco [19, 42]

OMEGA: I bet that TAU will now claim that programs cannot miscompute,[8] because they execute their program code and that is what they are supposed to do! The actual specification of Google Translate must have been "do a good and accurate translation".

TAU: I would be happy to say that programs cannot miscompute, but I accept that coding errors are a more general notion. Still, "do an accurate translation" is more of a marketing slogan than a specification.

TEACHER: Can we perhaps better understand the nature of the error if we consider what Google Translate team did in order to address it?

ALPHA: I can imagine two alternatives. If they had manual control over the training data, then they probably just removed the offending text from the dataset. If the training data is automatically collected, then they probably had to modify the behaviour of the scraping or training algorithm...

BETA: Now I can see why TAU says this was not an error. If we treat the training data just as one of the program inputs and the error was caused by inappropriate training data, then the error was neither *failure* nor a *mistake*. However, if Google had to modify their algorithm, then I would characterize the error as a *mistake*.

EPSILON: Are you suggesting that we can only decide whether the behaviour was a mistake if we know what Google did in order to address it? This is quite strange!

BETA: If we want to treat training data as separate from the program, then we need to think about the specification for the data too – training data is just too important in machine learning. Translating Russia as Mordor is an amusing story, but there are more dangerous biases that poorly chosen training dataset can cause!

EPSILON: I think that another useful lesson from the Google Translate case is that you never quite know when your specification is complete. As new unexpected situations arise, your specification has to evolve too. The fact that translating Russia as Mordor is not an "accurate translation" only becomes apparent once it happens. Interestingly, mathematicians went through exactly the same process when trying to define polyhedron![9]

TEACHER: It seems that the situation with mistakes is quite subtle. Now, can we at least agree on slips and failures? Or is it equally hard to recognize an incorrect encoding of a correct specification?

---

[7] As discussed by Northrop et al. [38], challenges of ultra-large-scale systems often need to be seen through this wider perspective.

[8] A discussion along these lines followed in response to Primiero and Fresco [19]; Dewhurst [15] argues that design errors should not be seen as miscomputation and Floridi et al. [18] discusses whether software can be seen as artifacts that can malfunction.

[9] EPSILON refers to the concept of *concept stretching* of Lakatos [31].

Omega: I would probably agree on slips, but the idea of failures makes an incorrect assumption that software can always have perfect and clear specification. A perceived failure can equally be a sign that you found an interesting complex scenario that you did not properly consider in your specification.

Alpha: In test-driven development, this would be the case where you add a new test case to clarify the specification. Strictly speaking, if you treat your tests as a specification, then you cannot have failures and all errors are either slips or mistakes. That is an interesting and for me quite unexpected consequence of the definition...

Omega: Another issue with our definition is that many interesting kinds of software cannot even have a clear specification. This is the case in live coded music or generative art where errors may turn out to be your most interesting pieces of work.[10] Outside of art, a data scientist or a data journalist trying to find an interesting story in a leaked data dump does not, at first, know what exactly she might be looking for!

## 2 Understanding programming paradigms through errors

As we have seen, the definition and classification of errors varies across different communities. This is also the case when dealing with errors. Not only do different groups follow different approaches to address errors, but their criteria for what can be considered a solution differ too. In philosophy of science, such sets of incommensurable assumptions are known as *research programmes* or *paradigms*.[11]

Thinking about coding errors gives us a new way of thinking about different programming paradigms and about the basic assumptions they do not question. While our class can reconcile some of the views (e.g. types and tests), we find that errors draw a bigger dividing line when trying to understand what a program is.

**Lesson 4: Maintenance as part of the programming process**

Teacher: According to some studies, from 1960s on, software maintenance has represented between 50% to 70% of total expenditure on software [17]. This is in part due to fixing bugs, but perhaps more importantly also due to required modification in response to the changes in the business environment. The Y2K bug is a good example of a change in the environment – the year change that was perhaps unexpected when software was originally written suddenly became important at the end of 1990s.

To open this lesson, is software maintenance something we have to do because we are unable to produce software without errors, or is it an inevitable part of the software development process?

---

[10] Some of the work by Anders Hoff collects interesting pieces produced accidentally by a buggy generative algorithm [25]

[11] This is an over-simplification, but for the purpose of this article, we use research programmes [30] and research paradigms [29] in a similar sense.

**Learning to live with errors**

ALPHA: The test-driven approach to software development supports both the initial development and maintenance. You write tests to specify the behaviour. With the Y2K bug, you would simply add a new test, fix possible failures and deploy a new version.

TEACHER: This is an interesting perspective on errors. Are you not only accepting that there are errors in the program, but even propose to incorporate them into the program development life-cycle?

TAU: Excuse me, but I think we should not be confusing what a program is with how it is created. A program is just a linguistic entity that you can analyse for correctness and run. How it is written is not a concern of our present discussion! Once we learn how to construct correct programs, no maintenance will be needed.

TEACHER: Treating program as a linguistic entity is exactly the kind of assumption that defines a scientific research programme...

OMEGA: And I think it is fundamentally flawed assumption! The fact that more than half of costs involve maintenance only supports this. The Y2K bug is an obvious issue in retrospect, but I can see how everyone missed it in 1980s. The environment in which software exists evolves and maintenance is more about adapting to the change than about fixing errors.

In other words, programming is not a task of constructing a linguistic entity, but rather a process of working interactively with the programming environment, using text simply as one possible interface.[12]

BETA: This reminds me of a banking system in Smalltalk that I worked on in the 1990s. In Smalltalk, you create software by interacting directly with the environment. The program runs at the same time as you are creating it. This was useful for rapid feedback, but once the system was working, you did not continue to live code it, except sometimes when it went wrong and you needed to investigate and fix the error.

OMEGA: But the decision when to watch the system and when to leave it alone was only yours! In live coded music, you want to interact with the system during the whole performance. In other applications, you interact more frequently in an early phase and less frequently in a later phase. But you still need to be able to interact!

**Lesson 5: Programming as human-computer interaction**

TEACHER: Let us spend more time on the idea of seeing programming as an interaction with the programming environment. We already talked about it in the context of live coded performance and Smalltalk programming. Those are great examples, but can we find more examples of this approach to programming?

OMEGA: Going back to the Y2K bug, live coding is actually what most companies did in response! Yes, they tried to fix possible bugs in advance, but in the end, they had a programmer on-site over the midnight to fix any problems as they happen.

---

[12] OMEGA is paraphrasing the definitions of Algol and Smalltalk research programmes [41]

Similarly, when mitigating a hacking attempt, you should be able to connect to the system and live code a defense [11]. Sadly, most systems just did not have a very good live coding interface to make such interaction possible.

Teacher: Is there anything that programming can learn from live coding then?

Epsilon: I have no interest in live coded music, but I see one similarity. What Omega described sounds like working with a REPL environment in Python or Racket.

There, you also type commands and run them immediately. If you get an error message, you can immediately correct yourself. REPL is no doubt "serious programming" but you still would not create a whole complex system just in REPL.

Teacher: Should we then agree that Omega's live coded music does not teach us much about professional software development, or are there other areas of software that are more similar to live coding?

Epsilon: I have one more example, but again, it is not what I normally think of as programming. Doing data science using tools like Matlab or R also feels like live coding. You also issue commands, observe the results and correct errors. Error like typos are immediately obvious, but you can also live code experiments to test more complex machine learning algorithms such as the one translating Russia as Mordor.

Tau: Following the same logic, you could claim that interactive theorem proving in Coq is also live coding with errors, because you do write code interactively until you satisfy your proof obligations. But unfinished proof is clearly not an error. Just a process of constructing a provably correct program.

Alpha: I'm not sure about theorem proving, but test-driven development (TDD) can probably be seen as a form of live-coding too. A good test runner runs in the background and shows the failing tests immediately as you are writing your code to provide rapid feedback.

Omega: This is exactly how live coding works! In TDD or data science, you are trying to make errors more visible so that you can quickly correct them. It is the same as live coded music where errors are immediately apparent. When you play a wrong note in a live coded performance, you will immediately hear that.

I think there are two main lessons for software development. First, we need to abolish this artificial distinction between a phase when software is created and a phase when it is autonomously running. As the costs of maintenance showed, programming is almost never done. Second, we need to code in a way that make the errors visible. Just like you can immediately hear a wrong note, you should be able to hear or see an incorrect execution of a program, perhaps in some built-in monitoring tools.

Tau: Well, now I see what you are trying to achieve, but it gives me a headache! If we wanted to guarantee correctness of such systems and account for all possible interactions, we would have to shift from proofs about static programs to proofs about interactions.

I'm not saying it is impossible. It might even be an interesting research problem! But I would very much prefer to solve the easier task of proving correctness of static programs first.

**Learning to live with errors**

**Lesson 6: Achieving correctness through tests and proofs**

TEACHER: Let us go back to the idea of programs as linguistic entities. Does this perspective help us eliminate coding errors?

TAU: Absolutely! A program is essentially a formal language term and this lets us utilize the resources of logic in order to increase the confidence in correctness of programs.[13] Instead of testing or debugging a program, one should prove that a program meets its specification. Then there will be no doubt that our software is correct and serves its purpose.

EPSILON: I like reading papers about programming language theory and I do enjoy an elegant proof, but I do not see how you could write proofs about software system that has hundreds of thousands of lines of code.

TAU: For reliable software engineering, we need to make proof an inherent part of the development process. Thanks to the Curry-Howard isomorphism, proofs are types. This means that all the amazing tools developed in logic are readily available in programming.

ALPHA: I can understand why people like types. They can be useful for avoiding basic kinds of errors. But the problem is that types do not capture the full complexity of software requirements.

You still need to write your user stories and to make sure they keep working, you need some form of testing. And if you will have tests anyway, why restrict the expressive power of a language with types and not use tests as the way of ensuring program correctness?

TAU: Substituting tests for proofs is never going to be enough. With tests, you are just showing the absence of certain errors, not proving your program correct.[14]

In modern dependently typed languages like Idris and Agda, you can express the full program specification just in terms of types. And then we will finally be able to write provably correct software!

TEACHER: I find it interesting that we are shifting the focus from program code to the properties that specify what the program does...

EPSILON: At first, I thought all this talk about proofs is pointless, but now that you mention properties, I think there might be something useful here.

Finding properties about your program is equally important if you are writing property-based tests with tools like QuickCheck [13], which then check that a property holds for partially randomly generated inputs. With random testing tools, you can focus more on finding as many useful properties as you can, rather than on writing a long-winded proof for every single one of them. This is a very efficient way for eradicating what we called *failures* in our earlier discussion.

---

[13] Historically, this position first appeared with the Algol language. To quote Priestley [41]:
One of the goals of the Algol research programme was to utilize the resources of logic to increase the confidence that it was possible to have in the correctness of a program. As McCarthy [35] had put it, "[i]nstead of debugging a program, one should prove that it meets its specifications, and this proof should be checked by a computer program".

[14] TAU is paraphrasing Dijkstra's famous quote: "Program testing can be used to show the presence of bugs, but never to show their absence." [16]

**Lesson 7: Specifying program properties**

TEACHER: Going back to the Y2K bug, the Knight Capital glitch and the Google Translate issue, can you think of properties that we can prove or check with random testing?

ALPHA: In case of banking system and Y2K, the test suite could include a check for a property that adding calculated interest to a savings account never decreases the total balance. The Y2K bug would be easily discovered and eliminated before it could happen!

TAU: Excuse me, but this is no way of finding theorems about programs! You just picked one random property that you believe should hold, but this is not a methodology for constructing provably correct software.

Instead, we need to start from small correct building blocks. We need to go back to the basics and define what a *date* is and what are the properties of functions operating on dates. Assuming $n$ is a function that returns the next date and $\geq$ is the ordering on dates, we want to prove *monotonicity* stating that $\forall d . n(d) \geq d$.

OMEGA: I wonder how this accounts for leap seconds... I'm not saying it cannot be done, but if we get stuck discussing monotonicity of dates, how are we ever going to develop something as complex as banking system? And isn't the specification going to become terribly complex?

TAU: Today, most people who write software assume that the costs of formal program verification outweigh the benefits and that fixing bugs later is cheaper than investing into correctness from the very start.[15]

I agree we still have a long way to go, but many research projects show that this can be done. For example, CompCert [32] is a verified C compiler with a machine-checked proof that the generated executable code behaves exactly as prescribed by the semantics of the source program.

TEACHER: Is there something that we can learn from CompCert about finding properties to prove about large systems?

TAU: Unlike the property about not decreasing total balance that ALPHA suggested earlier, the property that is proved in CompCert is a complete specification. The types specify that the compilation preserves semantics of programs between the formally specified source language and formally specified machine code.

---

[15] TAU is paraphrasing the introduction from a recent book on certified programming [12]

**Learning to live with errors**

This means that you can infer the correct implementation from the type.[16] In Coq, this did not happen automatically, but with recent development in Idris [9], you will just need to write a sufficiently detailed specification using types and the implementation will be inferred for you.

Epsilon: And the circle is closed! You are proposing to shift all the complexity of programming from writing a solution to writing a specification of the solution. I do not see how this is any safer.

In today's languages, we have to analyze the implementation and we invent new abstractions and constructs to make this easier. In your dependently typed future, we will have to analyze the equally complex specifications and we will presumably be inventing new abstractions and constructs to make this easier!

Omega: I have to admit, we came to an interesting point here. Is it easier to write a correct concrete implementation or a correct abstract specification?

I can see how writing the specification might be easier for C compiler. For a business application or any more creative use of computer, a specification in terms of properties will be harder to understand than a concrete implementation in a high-level language.

## 3 Mathematics, science, engineering and art

Through the discussion in the previous section, we discovered several ways of thinking about programs. We have also seen that if you take one of them as granted, it is hard to understand what proponents of other perspectives are saying – the different assumptions forming the basis of a different paradigm make other approaches illogical.

In this section, we try to relate the different ways of thinking about programs with different forms of human activity. It turns out that seeing programs as linguistic entities has much in common with mathematics and testing can learn from philosophy of science. Viewing a program as a long-running system is best seen as engineering and focusing on the interactions in programming can draw useful ideas from art.

**Lesson 8: Learning from philosophy of science and mathematics**

Teacher: Just like errors are an interesting topic for computing, so are experimental failures an interesting topic when discussing science. Can we learn something interesting about computing if we see errors as experimental failures?

Tau: A theory is scientific if it can be falsified,[17] but we can never prove it true. I see an obvious parallel with the test-driven development methodology. Tests can falsify a program, but they can never prove that it is correct. Testing does not increase our confidence about program correctness, just like a scientific experiment that does not fail is not showing that a scientific theory is correct.

---

[16] This is a common view in the dependently-typed community. There is no canonical reference making this exact point, but the work of McBride and McKinna [34] is close.

[17] Tau follows the Popperian view of scientific theories [40], which has been largely influential, but can be challenged from many perspectives [10]

Teacher: So, is there really no information in the fact that a scientific theory keeps escaping refutation? Or that a software system passes a growingly comprehensive test suite?

Epsilon: I think that even in science, not everybody agrees with the strict Popperian view. If a scientific theory passes a test and makes a correct prediction that could have disproved it, the probability that it is correct increases. Running the exact same test twice does not improve the probability, but I think that a new test does.

Alpha: I get it now! It's like in the Bayes' theorem. There is a prior probability that the software is correct, possibly based on the team behind it. Testing the software then provides new evidence and the Bayes' theorem prescribes how probabilities are to be changed in light of this new evidence.[18]

The fact that running the exact same test twice does not improve the probability of correctness, is already accounted for by the Bayes' theorem. So what we need is a way of finding relevant tests that do increase the probability.

Teacher: It seems that we can account for an increasing confidence in program correctness through testing by using the Bayesian approach to philosophy of science! Tau, do you agree?

Tau: That might be right, but what it shows is the limits of relying on tests. Perhaps you can increase probability, but you will never be absolutely certain. However, programming is more like mathematics where proofs can give you exactly that – absolute certainty. Experimentation is good in areas where we cannot have deductive knowledge,[19] but why settle for probability if you can have a proof?

Epsilon: Let me quote Imre Lakatos, who once said that "when a powerful new method emerges, the problems it cannot talk about are ignored" [31]. This is one danger of focusing on proofs – suddenly it becomes impossible to talk about important problems that cannot be explained through proofs. For example, what is the most intuitive way of modelling a distributed system? Surely, using the right intuitive programming model is crucial for writing correct software!

Teacher: Well, even if we see programming more as mathematics, is achieving correctness the only goal of proofs?

Omega: Proofs aren't there to convince you that something is true. They are there to show why it is true [4]. Even more interestingly, they sometimes play the same role as errors in live coding performance – they lead into unexplored territory of mathematics where we may even find different fundamental questions! [22]

Teacher: Can proofs about programs play this role?

---

[18] Bayesian epistemology of science [10] is based on similar reasoning

[19] As noted by Hacking [23], "we find prejudices in favour of theory as far back as there is institutionalized science."

**Learning to live with errors**

Epsilon: I had a look at some Coq programs and I am convinced that the proofs are true, but I doubt they fulfill the explanatory role. Reading a sequence of tactic invocations is definitely harder than reading an implementation of the same problem in a simple language.

Tau: I agree that Coq proofs can be complex, but they do have explanatory role. After all, in CompCert, the proof proceeds by a sequence of steps that translate source code in C into source code in a number of intermediary languages. Each of these translations preserves the semantics. The proof essentially describes the architecture of the compiler!

Omega: I can see how the CompCert proof reveals the problem decomposition, but I still doubt this approach can reveal truly new and unexplored ideas. Perhaps the problem is that constructive proofs like those about programs just have to follow more rigid structure. In mathematics, proof by contradiction can introduce unexpected creative twists that you can hardly get in proofs about programs.

**Lesson 9: Errors as a source of artistic creativity**

Teacher: Talking about sources of creative inspiration, Omega, you claimed earlier that errors are fun and can lead to new discoveries. Can you clarify what you mean?

Omega: An error in the performance of classical music occurs when the performer plays a note that is not written on the page. In genres that are not notated so closely, there are no wrong notes – only notes that are more or less appropriate [8]. A musical live coding performance is also not closely notated. You issue commands. Some of them are more appropriate and some of them are less appropriate.

Going back to our classification of errors, a slip such as syntax error rarely produces inspiring unexpected results. Failures and especially mistakes where you do not fully think through the consequences of your idea often lead to interesting creative moments. I imagine data scientists using REPL experience the same feelings when things do not go as expected!

Teacher: This is talking about more interactive flavor of programming, but do errors also have a useful role in more traditional software development?

Beta: I think so, but we need to consider the entire software development process, including clarification of the specification and maintenance. I believe that errors often show areas where we need to clarify specification.

Epsilon: In Erlang, we say that an error is a situation in which the programmer does not know how the program should correctly react, which seems to agree with what Beta is saying! Of course, in Erlang, the answer is to kill the process and let the supervisor recover from the error [3].

That way, an error caused by a rare interplay of multiple systems becomes just a *failure* that the system can automatically recover from, without turning into a *mistake* that programmers would need to immediately address.

Omega: I did not think of that, but I think this perfectly explains why artistic inspiration is important for programming! Seeing software development as mathematics or science makes us want to carefully control the process and impose tight constraints to make it reproducible. In contrast, artistic process tolerates or even welcomes a variety of inputs and works to produce the best result given the situations encountered while executing the process [21].

Tau: I'm not sure I follow. How is reproducibility a bad thing?

Omega: Of course, reproducibility is nice, but it is a chimera. There is much more variety in software engineering than in traditional engineering disciplines and requirements are always changing. An artistic process can adapt to very poorly stated requirements or even no requirements at all [21].

Beta: But Omega, do you not agree that the attempts to turn our discipline from what I earlier called "black art of programming" into an engineering discipline contributed to the quality of software that we build today?

Omega: It very likely did, but we can't say it's helped the quality of design much [21] and this is why we keep making the same errors!

In live coding you can very visibly improve through practice and the progress is clear to see. Perhaps we need to look at how artistic work makes such improvements possible? I suspect it might, in part, be thanks to the interaction with directly apparent errors and their unexpected consequences!

**Lesson 10: Engineering, or how did software get so reliable without proof?**

Teacher: The fact that we have this lecture series clearly shows that errors are still an important concern, but going back to what Beta said in the previous lecture, do you think that the quality of software that we build has been improving over, say, the last 60 years?

Epsilon: Looking at the history, it is fascinating to see all the regular warnings about the desperate state of the software industry. There is an interesting dichotomy: on the one hand, software is one of the largest and fastest-growing sectors of the U.S. economy; on the other hand, its reputation continues to be marked by perceptions of crisis and failure [17].

Tau: I'm often surprised how did software get so reliable without proofs.[20] Perhaps it is because solid engineering does help to limit the damage caused by broken software that we continue to produce.

That said, software is either correct, or not correct. Just like in mathematics you either have a proof or you do not, in software there is no such thing as "only slightly

---

[20] This very same question has been asked by a proponent of formal methods and the Algol research programme [24]

broken".[21] And so I agree with all of those who think that our industry has a serious problem to solve.

Beta: I disagree. The binary distinction between correct and incorrect software may be nice theory, but it does not reflect the reality. In practice, software might not be perfect, but can still be acceptable [44]. I believe this is what Epsilon is arguing for with Erlang's supervision model too, but the same ideas can matter when saving energy [39] or processing big data [37].

Epsilon: Absolutely! And even if we can make all the software components provably correct, we will still have the same problem – take large data centres or distributed systems in telecommunications as examples. At that scale, you need to account for unexpected hardware failures.

There will always be errors and letting the program crash and then recover automatically through supervision is an effective way of addressing that. However, I do like some of Omega's ideas about live coding and I think they can be used to make our monitoring and recovery tools better.

Teacher: It seems that there must be some useful lessons from other engineering disciplines, say, civil engineering...

Beta: The key idea in engineering is *safety factor*, which measures how much stronger a system is than it usually needs to be. A civil engineer will calculate the worst case load for a beam, but then make it ten times stronger. Such over-engineering is extremely effective and is even required by law for bridge building.

The idea of safety factors is something I would like to see in software engineering too. Perhaps not required by law, but certainly required by the code of ethics of any reputable software engineering organization [2].

Epsilon: I never thought about it in such a formal sense, so I have to admit, I do not even know how to calculate such safety factor for supervisor-based distributed applications...

Tau: Now you got me interested. The safety factor for a bridge can be calculated based on the expected load, but this assumes certain linearity. Increasing the strength of material twice provides a safety factor of 2. Software systems involve feedback loops or non-linearity where safety factor of 2 requires tenfold over-engineering.

What we need is a theory of stability, or perhaps a type system that can guarantee that certain amount of supervision does, indeed, provide the required safety factor!

Teacher: Judging by the way you are talking about the idea, Tau, it almost seems that Epsilon convinced you to accept errors! Or am I mistaken?

Tau: But we are not talking about errors here at all! All the so called "invalid states" that Epsilon proposes to handle by supervision and recovery are now perfectly valid and expected! It is just a different way of expressing your program, but as all the ones we talked about earlier, it does eliminate all errors from the software system.

---

[21] Tau is paraphrasing Erik Meijer's article [36]

## 4  What can software development learn through errors

Programming is not just one of mathematics, science, engineering or art. We can see it through all these four perspectives and each of them provides different inspiration. We have seen that each of the different paradigms for thinking about errors is more inclined to take inspiration from a different kind of human activity.

It might seem that these four approaches define incommensurable paradigms, but focusing our discussion around errors gives us a common central point that everyone can understand, even though in a different and somewhat incompatible way.[22]

In the last section, we will explore how thinking about errors gives us a new perspective on the world of software engineering. First, we consider how different programming styles align with different ways of thinking about errors. Second, we try to characterize computing problems based on what approach to errors is more appropriate.

**Lesson 11: The world according to errors**

TEACHER: Now that we discussed errors from many different perspectives in the last 10 lessons, perhaps the best way to start the last lesson is by naively repeating our initial question. What are errors and what can we do to build better software?

BETA: The simple answer is that correct software corresponds to its specification, but we have seen that there are many issues with this answer. A complete specification is hard to obtain, evolves during the life-cycle and for some systems, it may be very complex or even non-existent.

TAU: I think our earlier classification of different kinds of errors is useful here. I believe that formal methods can help us eliminate all failures and slips. Dealing with mistakes will be harder. They happen when we have gaps in our specification and, even in mathematics, you have to rely on intuition to guarantee that you are proving the right theorem.

EPSILON: I agree that dealing with failures is easier. Erlang programmers are used to living with failures and you could even say that we deliberately turn some mistakes into failures! As I mentioned, the mantra is "if you do not know what to do, let it crash!" This tells you to turn gaps in the specification, or mistakes, into failures, or crashes, that the supervisor process can deal with.

BETA: It is curious to see that Erlang encourages turning mistakes into failures. I always thought that in a poorly designed system, mistakes caused by unclear specification will be exhibited as technical failures.

---

[22] In sociology, the term boundary object [45] refers to objects that are understood by multiple communities and allow transfer of knowledge between them.

**Learning to live with errors**

This also suggests we may need to be careful about over-engineering. It works well for failures caused by genuine errors in encoding correct specification, but failures that are indication of design mistakes can rarely be fixed by, say, restarting.

ALPHA: It is interesting that you both are willing to accept failures. As I mentioned before, if we see tests in test-driven development as a specification, then there are no failures in TDD. The whole point of the methodology is to help us discover the specification and avoid mistakes.

EPSILON: I have seen way too many heated arguments between people arguing for types and people arguing for tests and now you are tell me that types and tests are about different kinds of errors? As TAU says, types help you eliminate failures and ALPHA now says that tests are mainly about discovering specification and avoiding mistakes! Perhaps this explains why many people use tests and types happily together.

OMEGA: I am not surprised. People are similarly confused about live coding and types. Live coded music or interactive data science can be perfectly well done in programming languages with types. Types help you avoid failures, but do not prevent you from making interesting mistakes that cause *defamiliarization* [20] which is a valuable source of creativity!

**Lesson 12: Learning to live with errors**

TEACHER: There is yet another way to look at the problem of errors. Would it be reasonable to accept some degree of errors in business applications and try to build fully specified and provably correct software only for mission critical systems?

BETA: This fits nicely with our earlier discussion about acceptability. A business application is acceptable even when you occasionally need to click some button twice, but acceptable medical software needs to work reliably.

EPSILON: We talked about the difficulties with finding properties to prove though. In fact, the traditional division between "mission critical" systems that must be formally verified and "business applications" where some errors are acceptable is misleading!

Instead, we should think about *property-based* systems whose behaviour is easier to describe as a property or a theorem and *behaviour-based* systems whose behaviour is better described as concrete code. For the first, we need to get the specification right and then we can use property-based testing or even dependently-typed languages that infers the implementation. For the latter, we need to build high-level languages that make the behavior they encode as easy to understand as possible.

TAU: I accept that formally specifying complex systems might be hard, but in case of mission critical systems it would be unethical to give up. We should try to prove at least some aspects correct, if only to increase our own understanding and confidence.

ALPHA: I can see how you can understand the system better if you try to create a minimal mathematical model, but didn't we agree that small models that ignore many details, like the representation of years, are never going to guarantee correctness?

TEACHER: Could we perhaps draw the dividing line between what we formalize and what we omit in a different way?

TAU: Good question. We could instead try to structure the system into multiple layers! A smaller low layer would then be formally verified. The higher layer could be built on top of the lower layer and so the properties guaranteed by the low layer would hold, even if the complexity of the higher layer made it hard to make any formal statements about that part.

OMEGA: In general, I agree that a lower layer needs to be more correct than higher layer. But even with a proof, I think we cannot get it absolutely right[23] and we need other mechanisms to address potential errors. Sometimes, a supervision model like that advocated by EPSILON might be good enough, but ultimately, you will always need manual intervention in some cases.

TAU: Are you suggesting that programmers need to be readily available and watch their program all the time? Surely, nobody can actually work in this way!

OMEGA: Quite the opposite. The DevOps movement [26] is, in many ways, doing exactly that! It encourages close collaboration between development and operations teams to enable automation and frequent releases.

You can see that as live coding the whole deployment environment, albeit with very poor tools! Rather than having programming environments that have been designed to support manual intervention like Smalltalk, DevOps teams rely on ad-hoc monitoring tools and have limited ways for direct interactions. We need to finally accept that software is a living system and build tools accordingly. I suppose this goes back to not treating programs as just linguistic entities...

TAU: There is one thing I still do not understand about the idea of live intervention though. How can you react quickly enough? In the Knight Capital glitch, it took 45 minutes and perhaps you could live code a fix in minutes, but is that ever going to be fast enough?

Using your favourite analogy with music – a manual intervention, like a guitarist lifting his finger from a discordant note [8] will have reaction time a fraction of a second. Significantly shorter than any live coder can ever hope...

OMEGA: Yes, that is true, but live coding does not mean just typing code so that you fix an error. You can also live code your environment, to make quick reactions easier. A quick intervention then gives you enough time to think and decide on the best course of action. In case of live coded music, we do this using flexible interfaces like emacs and monome,[24] but there is no reason why the same ideas would not work in other environments.

EPSILON: Quite intriguing! This idea of fast and slow ways of reacting reminds me of Kahneman's two modes of thought [27]. The fast system is automatic, instinctive and

---

[23] A recent work [28] discovered errors in nine out of nine academic papers presenting a formal model, two of which were mechanized.
[24] The idea is inspired by the performance of the Meta-eX group [1]

**Learning to live with errors**

emotional while the slow system is more effortful, deliberative and logical. Through deliberate training and practice, you can train your fast system to react differently. So perhaps our new model is even getting close to how the human thought works!

## 5 Summary: Escaping the crisis narrative

In this paper, we discussed a wide range of ideas about programming through the perspective of program errors or miscomputations. This point of view provides a new way of defining programming paradigms and thinking about software more generally. Because of the ubiquitous crisis narrative that is persistent in our industry, this point of view might be more revealing than when considering traditional programming paradigms such as functional or object-oriented programming.

**Acknowledgements** This work was supported by The Alan Turing Institute under the EPSRC grant EP/N510129/1. I would like to thank to fellow members of the Revolution & Beers Cambridge group, namely Dominic Orchard, Sam Aaron, Antranig Basman and Stephen Kell.

The characters in this essay are not entirely fictitious and I would like to thank (and apologize) to colleagues who inspired them. Last but not least, I'm grateful to anonymous reviewers for thought provoking feedback that inspired many improvements in the essay. Any remaining errors are mine and I'm willing to live with them.

## About the author

**Tomas Petricek** Tomas is a Visiting Researcher at the Alan Turing institute, working on tools for open data-driven storytelling (http://thegamma.net). His many other interests include programming language theory (his PhD thesis is on *coeffects*, a theory of context-aware programming languages), open-source and functional programming (he is an active contributor to the F# ecosystem), but also understanding programming through the perspective of philosophy of science. Contact him at tomas@tomasp.net.